\documentclass[]{aa}

\usepackage{graphicx}
\usepackage{natbib}
\bibpunct{(}{)}{,}{a}{}{;}
\usepackage{amssymb}
\usepackage{amsmath}
\usepackage{url}
\usepackage{multirow}
\usepackage{txfonts}

\begin{document}

\title{Observational evidence for dissociative shocks in the inner 100 AU of low-mass protostars using \textit{Herschel}-HIFI\thanks{\textit{Herschel} is an ESA space observatory with science instruments provided by European-led Principal Investigator consortia and with important participation from NASA.}}
\titlerunning{Observational evidence for dissociative shocks in the inner 100 AU of low-mass protostars}

\author{L.E. Kristensen\inst{1,2}
\and E.F. van Dishoeck\inst{1,3}
\and A.O. Benz\inst{4}
\and S. Bruderer$^3$
\and R. Visser\inst{5}
\and S.F. Wampfler\inst{6,7}
}

\institute{
%1
Leiden Observatory, Leiden University, PO Box 9513, 2300 RA Leiden, the Netherlands \and
% 2
Harvard-Smithsonian Center for Astrophysics, 60 Garden Street, Cambridge, MA 02138, USA\\ \email{lkristensen@cfa.harvard.edu} \and
% 3
Max Planck Institut f{\"u}r Extraterrestrische Physik, Giessenbachstrasse 1, 85748 Garching, Germany \and
% 4
Institute for Astronomy, ETH Zurich, 8093 Zurich, Switzerland \and
% 5
Department of Astronomy, University of Michigan, 500 Church Street, Ann Arbor, MI 48109-1042, USA \and
% 6
Centre for Star and Planet Formation, Natural History Museum of Denmark, University of Copenhagen, {\O}ster Voldgade 5-7, DK-1350 Copenhagen K, Denmark \and
% 7
Niels Bohr Institute, University of Copenhagen, Juliane Maries Vej 30, DK-2100 Copenhagen {\O}, Denmark
}

\date{\center{Received 1 April 2013 / Accepted 24 June 2013}}

%%%%%%%%%%%%%%
% BEGIN TEXT %
%%%%%%%%%%%%%%

\abstract
{}
{\textit{Herschel}-HIFI spectra of H$_2$O towards low-mass protostars show a distinct velocity component not seen in observations from the ground of CO or other species. The aim is to characterise this component in terms of excitation conditions and physical origin.}
{A velocity component with an offset of $\sim$ 10 km s$^{-1}$ detected in spectra of the H$_2$O 1$_{10}$--1$_{01}$ 557 GHz transition towards six low-mass protostars in the `Water in star-forming regions with \textit{Herschel}'  (WISH) programme is also seen in higher-excited H$_2$O lines. The emission from this component is quantified and local excitation conditions are inferred using 1D slab models. Data are compared to observations of hydrides (high-$J$ CO, OH$^+$, CH$^+$, C$^+$, OH) where the same component is uniquely detected.}
{The velocity component is detected in all six targeted H$_2$O transitions ($E_{\rm up}$ $\sim$ 50--250 K), as well as in CO 16--15 towards one source, Ser SMM1. Inferred excitation conditions imply that the emission arises in dense ($n$\,$\sim$\,5$\times$10$^6$--10$^8$~cm$^{-3}$) and hot ($T$\,$\sim$\,750~K) gas. The H$_2$O and CO column densities are $\gtrsim$\,10$^{16}$ and 10$^{18}$~cm$^{-2}$, respectively, implying a low H$_2$O abundance of $\sim$\,10$^{-2}$ with respect to CO. The high column densities of ions such as OH$^+$ and CH$^+$ (both $\gtrsim$\,10$^{13}$~cm$^{-2}$) indicate an origin close to the protostar where the UV field is strong enough that these species are abundant. The estimated radius of the emitting region is 100~AU. This component likely arises in dissociative shocks close to the protostar, an interpretation corroborated by a comparison with models of such shocks. Furthermore, one of the sources, IRAS4A, shows temporal variability in the offset component over a period of two years which is expected from shocks in dense media. High-$J$ CO gas detected with \textit{Herschel}-PACS with $T_{\rm rot}$\,$\sim$\,700~K is identified as arising in the same component and traces the part of the shock where H$_2$ reforms. Thus, H$_2$O reveals new dynamical components, even on small spatial scales in low-mass protostars.}
{}

\keywords{Astrochemistry --- Stars: formation --- ISM: molecules --- ISM: jets and outflows}

\maketitle

\section{Introduction}

Star formation is a violent process, even in low-mass protostars ($L_{\rm bol}$ $<$ 100 L$_\odot$). X-rays and UV radiation from the accreting star-disk system illuminate the inner, dense envelope while the protostellar jet and wind impinge on the same inner envelope, causing shocks into dense gas. The implication is that the physical and chemical conditions along the outflow cavities are significantly different from the conditions in the bulk of the collapsing envelope. Very little is known about the hot ($T$ $>$ 500 K) gas in low-mass protostars, primarily because the mass of the hot gas is at most a few \% of that of the envelope. Second, few unique, abundant tracers of the hot gas in the inner envelope exist, save H$_2$ and CO observed at near-infrared wavelengths \citep{herczeg11}, both of which are very difficult to detect towards the deeply embedded protostars where the $A_{\rm V}$ is $\gtrsim$100 \citep[e.g.,][]{maret09}.

The hot gas is most prominently seen in high-$J$ CO observations with the Photodetector Array Camera and Spectrometer (PACS) on \textit{Herschel} \citep{poglitsch10, pilbratt10}, where CO emission up to $J$ = 49--48 is detected towards low-mass protostars \citep[$E_{\rm up}$ $\sim$ 5000 K;][]{herczeg12, gecco12}. The high-$J$ CO emission ($J_{\rm up}$ $>$ 14) traces two components with rotational temperatures of 300 and 700--800~K (a warm and hot component, respectively) seen towards several tens of low-mass protostars \citep{green13, karska13, manoj12}. At present it is unclear whether the two temperature components correspond to separate physical components, or whether they are part of a distribution of temperatures, or even just a single temperature and density \citep{visser12, neufeld12, manoj12, karska13}. Moreover, depending on the excitation conditions, the rotational temperature may or may not be identical to the kinetic gas temperature. \citet{santangelo13} and \citet{dionatos13} relate the two rotational-temperature components seen in CO to two rotational-temperature components seen in H$_2$ rotational diagrams; with its lower critical density ($\sim$ 10$^3$ cm$^{-3}$ versus $>$ 10$^5$~cm$^{-3}$ for high-$J$ CO transitions) H$_2$ is more likely to be thermally excited suggesting that the excitation is thermal. 

The PACS lines provide little information, beyond the rotational temperature, as they are all velocity-unresolved and no information is therefore available on the kinematics of this gas.  If the very high-$J$ CO emission is caused by shocks in the inner dense envelope ($n$ $\gtrsim$ 10$^6$ cm$^{-3}$) the rotational temperature is similar to the kinetic gas temperature, as proposed by, e.g., \citet{vankempen10} and \citet{visser12}. Further support for this hypothesis and the existence of shocks in the dense inner envelope comes from observations of [\ion{O}{i}] at 63 $\mu$m and OH, also done with PACS. Towards the low-mass protostar HH46, the inferred column densities of these species indicate the presence of fast ($\varv$ $>$ 60 km s$^{-1}$) dissociative shocks close to the protostar \citep{vankempen10, wampfler10, wampfler13}.

As part of the `Water in star-forming regions with \textit{Herschel}' programme \citep[WISH\footnote{\url{http://www.strw.leidenuniv.nl/WISH}};][]{vandishoeck11}, \citet{, kristensen10, kristensen12} detected a distinct velocity component towards six low-mass protostars in the H$_2$O 1$_{10}$--1$_{01}$ 557 GHz transition with the Heterodyne Instrument for the Far-Infrared on \textit{Herschel} \citep[HIFI;][]{degraauw10}. The component is typically blue-shifted from the source velocity by $\sim$2--10 km s$^{-1}$ and the width is in the same range as the offset, $\sim$5--10 km s$^{-1}$. \citet{kristensen12} referred to this component as the ``medium component'' and associated it with shocks on the inner envelope/cavity wall based on the coincidence of H$_2$O masers and this velocity component. The maser association suggests excitation conditions where the density is $>$ 10$^7$ cm$^{-3}$ and $T$ $>$ 500 K \citep{elitzur92}, conditions similar to those inferred for the high-$J$ CO emission. Yet, so far little is known concerning this velocity component, its origin in the protostellar system and the local conditions, primarily because the component is not seen in ground-based observations of lower-excited lines towards these same sources.

In this paper, we combine observations of the H$_2$O offset component presented above with observations of light hydrides \citep[OH, OH$^+$, CH$^+$;][Benz et al. in prep.]{wampfler13} and highly excited velocity-resolved CO (up to $J$ = 16--15, for one source) to constrain the physical and chemical conditions in this velocity component. The component considered in this paper is the same as presented in \citet{kristensen12}, except towards one source, NGC1333-IRAS2A. The observations and data reduction are described in Sect. 2. The results are presented in Sect. 3. In Sect. 4 we discuss the derived excitation conditions and the interpretation of the velocity component in the context of irradiated shocks. Finally, Sect. 5 contains the conclusions.

\section{Observations}
\label{sec:obs}

\subsection{Source sample}

Six low-mass protostars in NGC1333 and Serpens clearly show the presence of an offset component in the H$_2$O 557 GHz line \citep{kristensen12}. The sources are NGC1333-IRAS2A, IRAS3A, IRAS4A, IRAS4B and Serpens SMM1, SMM3. Two sources, IRAS3A and SMM3, show the offset component in absorption against the outflow and/or continuum. The sources are all part of the WISH sample of low-mass protostars \citep{vandishoeck11, kristensen12}. IRAS3A was only observed in the H$_2$O 557 GHz line and is therefore excluded from further analysis. The component is not seen towards any other source, and the possible reasons will be discussed in Sect. \ref{sec:origin}. 

\citet{kristensen10, kristensen12} identified three characteristic components from the line profiles of H$_2$O lines in low-mass protostars: a narrow ($FWHM$ $<$ 5 km s$^{-1}$), medium (5 $<$ $FWHM$ $<$ 20 km s$^{-1}$), and broad ($FWHM$ $>$ 20 km s$^{-1}$) component. The component in this study is characterised by its offset rather than line width as in our previous work, and thus named ``offset'' component. For most sources, it is the same as the medium component, except for IRAS 2A. Towards this source the ``offset'' component is broader than the ``medium'' component (40~km~s$^{-1}$ vs. 10 km s$^{-1}$) but is clearly offset from the source velocity. The offset component is not seen in low-$J$ CO transitions from the ground but was detected in H$_2$O emission \citep{kristensen10}, which is not the case for the ``medium'' component, and hence the redefinition.

\begin{figure}[!t]
\begin{center}
\includegraphics[width=.98\columnwidth]{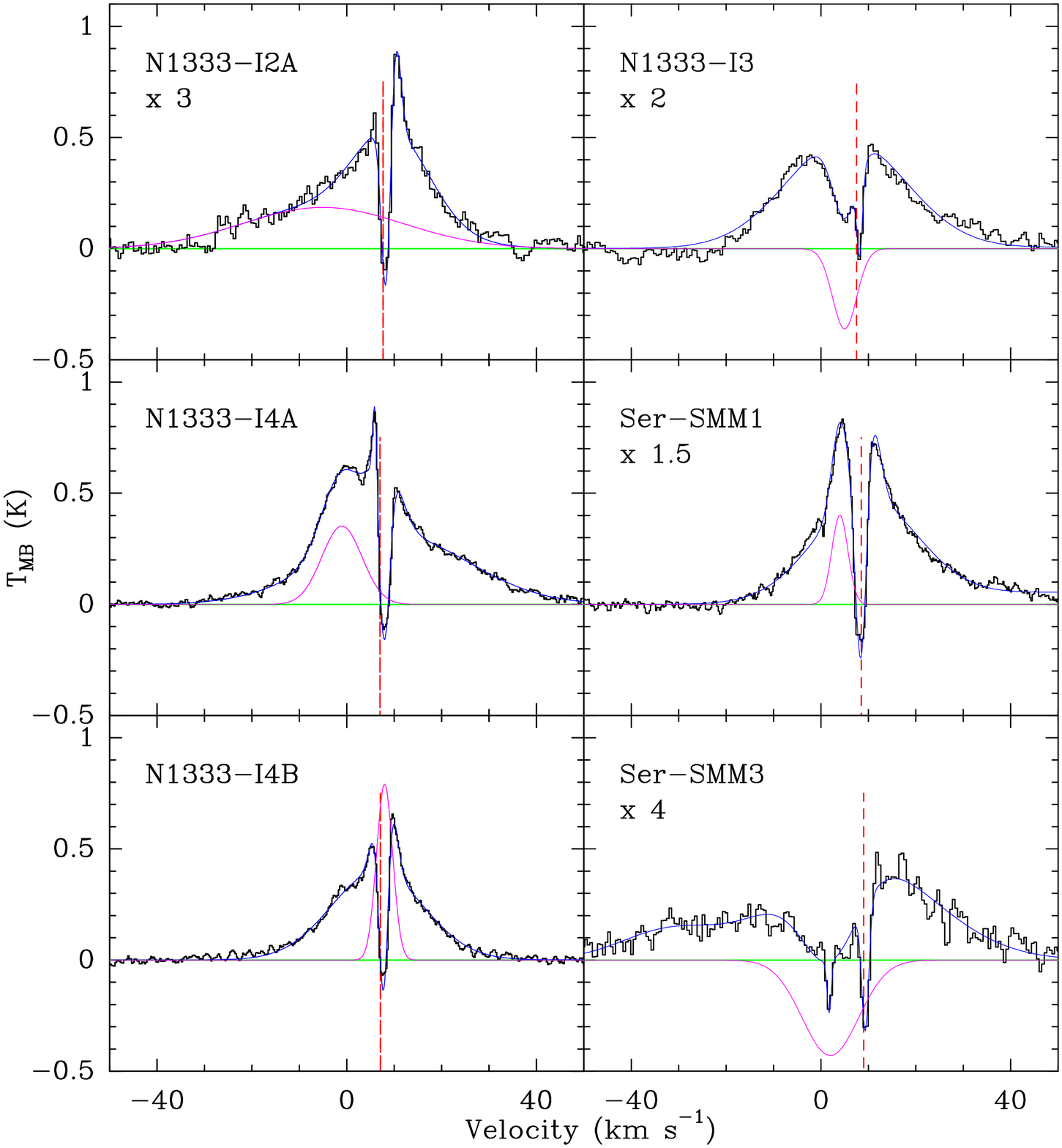}
\includegraphics[width=.98\columnwidth]{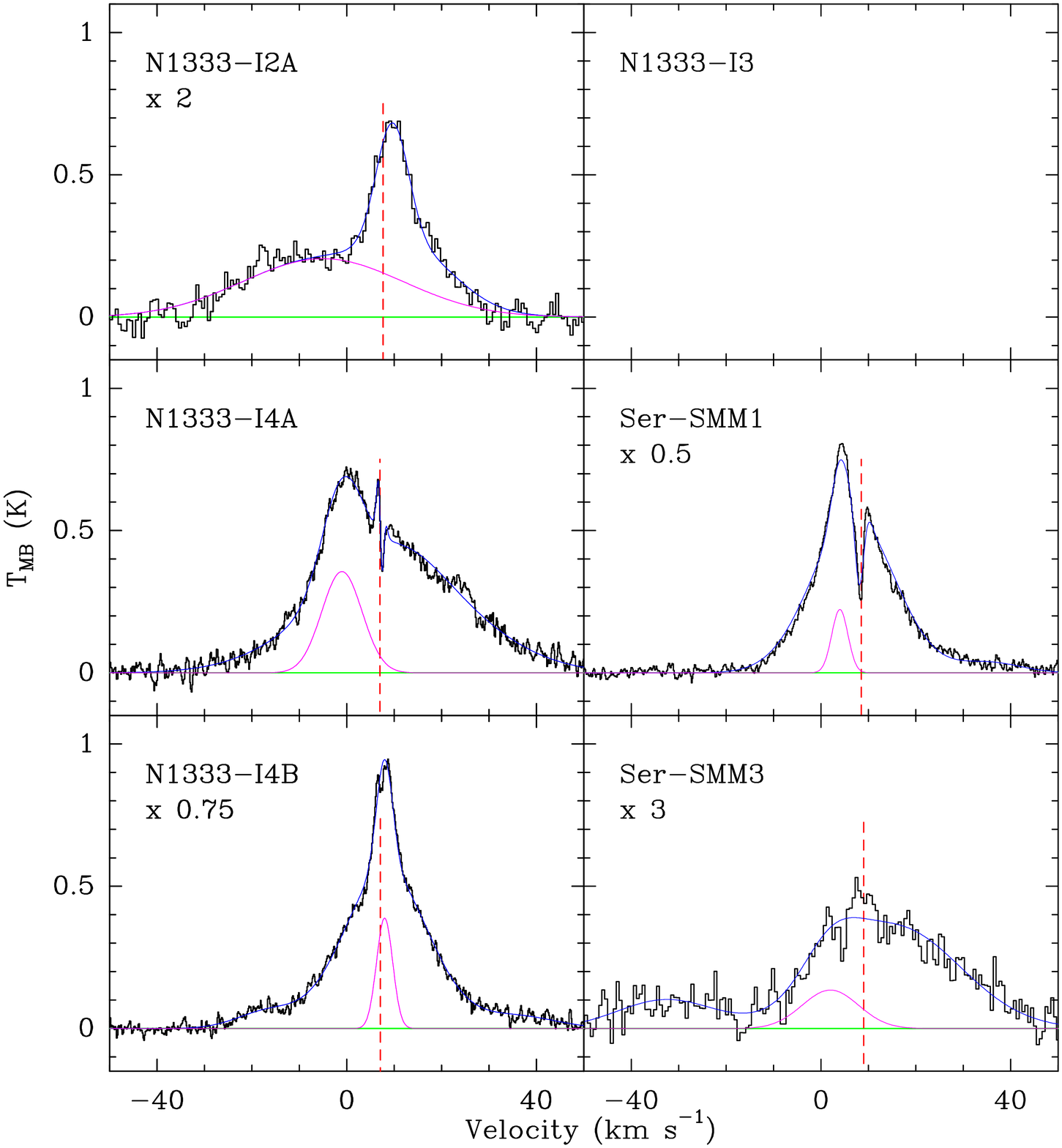}
\end{center}
\caption{\textit{Top:} Continuum-subtracted HIFI spectra of the H$_2$O 1$_{10}$--1$_{01}$ ground-state transition at 557 GHz ($E_{\rm up}$ = 60 K). The profiles have been decomposed into Gaussian components and the best-fit profile is shown in blue. The offset component is highlighted in magenta for clarity. The baseline is shown in green and the source velocity with a red dashed line. \textit{Bottom:} Same as the top figure, except that the line shown is the excited H$_2$O 2$_{02}$--1$_{11}$ line at 988 GHz ($E_{\rm up}$ = 100 K). More details on the decomposition are found in \citet{kristensen10, kristensen12} for the NGC1333 sources and Ser SMM3.}
\label{fig:spectra_all}
\end{figure}

\subsection{Herschel observations}

The central positions of the six low-mass protostars were observed with HIFI on \textit{Herschel} in twelve different settings covering six H$_2^{16}$O and two CO transitions, and HCO$^+$, OH, CH$^+$, OH$^+$ and C$^+$ ($E_{\rm u}/k_{\rm B}\approx50-300$ K; see Table \ref{tab:settings} for an overview). Only Ser-SMM1 was observed in all settings, the other sources in a sub-set (Tables \ref{tab:settings} and \ref{tab:obsid}).

Data were obtained using the dual beam-switch mode with a nod of 3\arcmin\ and a fast chop and continuum optimisation, except for the ground-state ortho-H$_2$O line at 557 GHz, where a position switch was used \citep[see][ for details]{kristensen12}. The diffraction-limited beam size ranges from 12\arcsec\ to 39\arcsec\ (2800--9200 AU for a distance of 235 pc). Data were reduced using HIPE ver. 8. The calibration uncertainty is taken to be 10\% for lines observed in Bands 1, 2, and 5 while it is 30\% in Band 4 \citep{roelfsema12}. The pointing accuracy is $\sim$2\arcsec. A main-beam efficiency of 0.65--0.75 is adopted (Table \ref{tab:settings}). Subsequent analysis of the data is performed in CLASS\footnote{\url{http://www.iram.fr/IRAMFR/GILDAS/}} including subtraction of linear baselines. H- and V-polarisations are co-added after inspection; no significant differences are found between the two data sets.

To compare observations done with different beam sizes, all components are assumed to arise in an unresolved physical component even in the smallest beams (12\arcsec). The emission is scaled to a common beam-size of 20\arcsec\ (the beam at 1 THz) using a simple geometrical scaling for a point source. We argue \textit{a posteriori} that this is an appropriate scaling (Sect. \ref{sec:excitation}).

\begin{table*}
\caption{Species and transitions observed with \textit{Herschel}-HIFI containing the offset component\tablefootmark{a}.}
\tiny
\begin{center}
\begin{tabular}{r c c c c c c c l}
\hline \hline
\multicolumn{1}{c}{Transition} & $\nu$ & $\lambda$   & $E_{\rm u}/k_{\rm B}$ & $A$  & Beam\tablefootmark{b} & $t_{\rm int}$\tablefootmark{c} & $\eta_{\rm MB}$\tablefootmark{b} & Sources \\
                               & (GHz) & ($\mu$m)    & (K)   & (s$^{-1}$) & ($''$) & (min.) \\
\hline
H$_2$O ~~~1$_{10}$--1$_{01}$ & \phantom{1}556.94 & 538.29 & \phantom{1}61.0 & 3.46(--3) & 38.1 & 13.0 & 0.75 & IRAS2A, IRAS3A, IRAS4A, IRAS4B, SMM1, SMM3 \\
          2$_{12}$--1$_{01}$ &           1669.90 & 179.53 & 114.4           & 5.59(--2) & 12.7 & 23.7 & 0.71 & IRAS2A, IRAS4A, IRAS4B, SMM1, SMM3 \\
1$_{11}$--0$_{00}$           &           1113.34 & 269.27 & \phantom{1}53.4 & 1.84(--2) & 19.0 & 43.5 & 0.74 & IRAS2A, IRAS4A, IRAS4B, SMM1, SMM3 \\
2$_{02}$--1$_{11}$           & \phantom{1}987.93 & 303.46 & 100.8           & 5.84(--3) & 21.5 & 23.3 & 0.74 & IRAS2A, IRAS4A, IRAS4B, SMM1, SMM3 \\
2$_{11}$--2$_{02}$           & \phantom{1}752.03 & 398.64 & 136.9           & 7.06(--3) & 28.2 & 18.4 & 0.75 & IRAS2A, IRAS4A, IRAS4B, SMM1, SMM3 \\
3$_{12}$--3$_{03}$           &           1097.37 & 273.19 & 249.4           & 1.65(--2) & 19.7 & 32.5 & 0.74 & IRAS2A, IRAS4A, IRAS4B, SMM1, SMM3 \\
3$_{12}$--2$_{21}$           &           1153.13 & 259.98 & 249.4           & 2.63(--3) & 18.4 & 13.0 & 0.64 & IRAS2A, IRAS4A, IRAS4B, SMM1, SMM3 \\
\hline
CO        ~~~ 10--9 &  1151.99 & 260.24 &           304.1 & 1.00(--4) & 18.4 & 13.0 & 0.64 & IRAS2A, IRAS4A, IRAS4B, SMM1, SMM3 \\
16--15              &  1841.35 & 162.81 &           751.7 & 4.05(--4) & 11.5 & 44.6 & 0.70 & SMM1 \\
\hline
CH$^+$  ~~~ 1--0 &  \phantom{1}835.14 & 358.97 & \phantom{1}40.1 &  2.3(--3) & 25.4 & 15.6 & 0.75 & IRAS2A, IRAS4A, IRAS4B, SMM1 \\
OH$^+$  ~~~ 1--0\tablefootmark{d} &            1033.12 & 290.18 & \phantom{1}49.6 &  1.8(--2) & 20.5 & 30.1 & 0.74 & IRAS2A, IRAS4A, IRAS4B, SMM1 \\
C$^+$   ~~~ 2--1 &            1900.54 & 157.74 & \phantom{1}91.2 & 2.30(--6) & 11.2 & 15.8 & 0.69 & IRAS2A, IRAS4A, IRAS4B, SMM1 \\
HCO$^+$ ~~~ 6--5 &  \phantom{1}535.06 & 560.30 & \phantom{1}89.9 & 1.27(--2) & 39.6 & 43.7 & 0.75 & IRAS2A, IRAS4A, IRAS4B, SMM1, SMM3 \\
CH ~~~ \phantom{6--5}\tablefootmark{d} &  \phantom{1}536.76 & 558.52 & \phantom{1}25.8 & 6.80(--4) & 39.6 & 43.7 & 0.75 & IRAS2A, IRAS4A, IRAS4B, SMM1, SMM3 \\
OH ~~~ \phantom{6--5}\tablefootmark{d} &            1834.75 & 163.40 &      269.8 & 2.12(--2) & 11.6 & 44.6 & 0.70 & SMM1 \\
\hline
\end{tabular}
\tablefoot{For lines with hyperfine splitting (OH and OH$^+$) only the strongest component is shown here.
	\tablefoottext{a}{From the JPL database of molecular spectroscopy \citep{pickett98}.}
	\tablefoottext{b}{Half-power beam width, from \citet{roelfsema12}.}
	\tablefoottext{c}{Total on $+$ off integration time incl. overheads.}
	\tablefoottext{d}{Transitions with hyperfine splitting.}}
\end{center}
\label{tab:settings}
\end{table*}

\begin{table*}
\caption{H$_2$O emission in the offset component.}
\tiny
\begin{center}
\begin{tabular}{l c c c @{} c c c @{} c c c @{} c c c @{} c c c @{} c c c @{} c c c @{} c c c}
\hline \hline
 & & \multicolumn{2}{c}{IRAS2A} && \multicolumn{2}{c}{IRAS3A} && \multicolumn{2}{c}{IRAS4A} && \multicolumn{2}{c}{IRAS4B} && \multicolumn{2}{c}{Ser-SMM1} && \multicolumn{2}{c}{Ser-SMM3} \vspace{1pt} \\ 
 \cline{3-4} \cline{6-7} \cline{9-10} \cline{12-13} \cline{15-16} \cline{18-19}
Transition & rms\tablefootmark{a} & $\int T_{\rm MB}$ d$\varv$ & $T_{\rm peak}$ && $\int T_{\rm MB}$ d$\varv$ & $T_{\rm peak}$ && $\int T_{\rm MB}$ d$\varv$ & $T_{\rm peak}$ && $\int T_{\rm MB}$ d$\varv$ & $T_{\rm peak}$ && $\int T_{\rm MB}$ d$\varv$ & $T_{\rm peak}$ && $\int T_{\rm MB}$ d$\varv$ & $T_{\rm peak}$ \\
  & (mK) & (K\,km\,s$^{-1}$) & (K) && (K\,km\,s$^{-1}$) & (K) && (K\,km\,s$^{-1}$) & (K) && (K\,km\,s$^{-1}$) & (K) && (K\,km\,s$^{-1}$) & (K) && (K\,km\,s$^{-1}$) & (K) \\ \hline
1$_{10}$--1$_{01}$ & \phantom{1}7 & 2.55 & 0.06 && --1.25 & --0.19 && 4.23 & 0.40 && \phantom{1}3.78 & 0.89 && 0.71 & 0.17 && --1.59 & --0.11  \\
2$_{12}$--1$_{01}$ &  80 & 7.73 & 0.17 && \ldots & \ldots && 2.40 & 0.23 && 16.35 & 3.84 && 4.51 & 1.06 && --2.08 & --0.14  \\
1$_{11}$--0$_{00}$ & 16 & 3.96 & 0.09 && \ldots & \ldots && 3.33 & 0.31 && \phantom{1}5.78 & 1.36 && 1.51 & 0.36 && --1.56 & --0.10  \\
2$_{02}$--1$_{11}$ & 16 & 4.65 & 0.11 && \ldots & \ldots && 3.67 & 0.35 && \phantom{1}2.17 & 0.51 && 2.54 & 0.60 && \phantom{--}0.69 & \phantom{--}0.05  \\
2$_{11}$--2$_{02}$ & 18 & 2.42 & 0.06 && \ldots & \ldots && 1.96 & 0.18 && \phantom{1}1.44 & 0.34 && 1.37 & 0.32 && \phantom{--}0.52 & \phantom{--}0.04 \\
3$_{12}$--3$_{03}$ & 56 & 3.14 & 0.07 && \ldots & \ldots && 1.17 & 0.11 && \phantom{1}1.11 & 0.26 && 1.66 & 0.39 && \phantom{--}0.25 & \phantom{--}0.02 \\
3$_{12}$--2$_{21}$ & 70 & 3.16 & 0.07 && \ldots & \ldots && 2.09 & 0.20 && \phantom{1}1.16 & 0.27 && 2.66 & 0.63 && $<$0.05 & $<$0.01 \\
\hline
$\Delta\varv$ (km\,s$^{-1}$) & & \multicolumn{2}{c}{40} && \multicolumn{2}{c}{6} && \multicolumn{2}{c}{10} && \multicolumn{2}{c}{4} && \multicolumn{2}{c}{4} && \multicolumn{2}{c}{14} \\
$\varv_{\rm LSR}$ (km\,s$^{-1}$) & & \multicolumn{2}{c}{--5} && \multicolumn{2}{c}{5} && \multicolumn{2}{c}{--1} && \multicolumn{2}{c}{8} && \multicolumn{2}{c}{4} && \multicolumn{2}{c}{\phantom{1}2} \\
$\varv_{\rm offset}$\tablefootmark{b} (km\,s$^{-1}$) & & \multicolumn{2}{c}{--12.7} && \multicolumn{2}{c}{--3.3} && \multicolumn{2}{c}{--8.0} && \multicolumn{2}{c}{0.9} && \multicolumn{2}{c}{--4.5} && \multicolumn{2}{c}{--5.6} \\ \hline
\end{tabular}
\tablefoot{Obtained from Gaussian fits to each component; negative values are for absorption. Upper limits are 1$\sigma$.
	\tablefoottext{a}{Measured in 1 km\,s$^{-1}$ channels.}
	\tablefoottext{b}{Offset velocity with respect to the source velocity as reported by \citet{yildiz13}.}}
\end{center}
\label{tab:intensity}
\end{table*}

\begin{figure}
\begin{center}
\includegraphics[width=\columnwidth, angle=0]{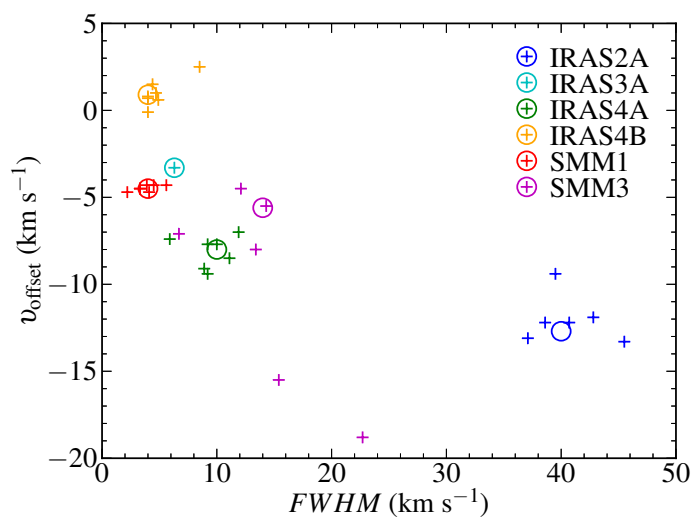}
\end{center}
\caption{Velocity of the offset component with respect to the source velocity as a function of $FWHM$ for all observed transitions. The plus signs show the results of the decomposition of each line, whereas the circles show the values of the offset and width chosen as fixed.}
\label{fig:vlsr_fwhm}
\end{figure}

\begin{figure}
\begin{center}
\includegraphics[width=0.75\columnwidth, angle=0]{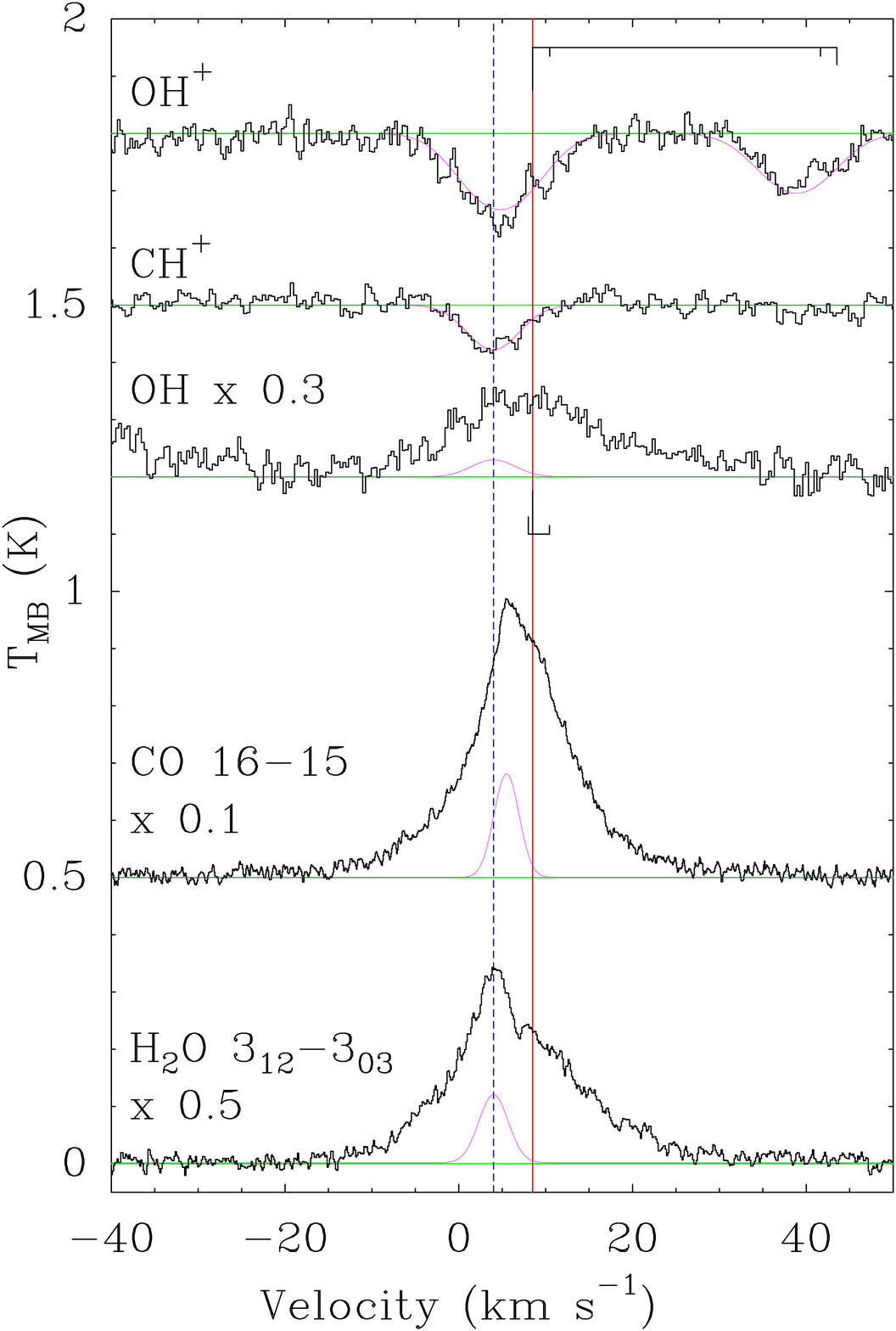}
\end{center}
\caption{Continuum-subtracted HIFI spectra of H$_2$O, CO, OH, CH$^+$, OH$^+$ obtained towards the central position of Ser SMM1 ($\varv_{\rm source}$ = 8.5 km s$^{-1}$). The red vertical line indicates $\varv_{\rm source}$, while the blue dashed line shows the position of the offset component. The blended OH triplet is centred on the strongest hyperfine component as indicated by the three vertical black lines situated directly beneath the OH spectrum. The same is the case for the OH$^+$ spectrum, with the location of the hyperfine components and their relative strengths indicated by the black lines above the spectrum. The spectra have been shifted vertically for clarity and in some cases scaled by a factor, indicated on the figure. CH$^+$ and OH$^+$ are fitted by a single Gaussian, OH by two, and CO and H$_2$O by three; the offset components are shown in magenta.}
\label{fig:spectra}
\end{figure}

\section{Results}
\label{sec:results}

\subsection{Quantifying emission from the offset component}\label{sect:gauss}

Figure \ref{fig:spectra_all} shows the H$_2$O 1$_{10}$--1$_{01}$ (557 GHz) and 2$_{02}$--1$_{11}$ (988 GHz) spectra towards all sources with an offset component; the offset component is marked in the figure. The remaining parts of each profile will be presented and analysed in a forthcoming paper (Mottram et al. in prep.). The component is characterised by being significantly blue-shifted ($\varv_{\rm source} - \varv_{\rm LSR} > 2$ km s$^{-1}$), or, for the isolated case of IRAS4B, by being located close to the source velocity ($\varv_{\rm source} - \varv_{\rm LSR} < 1$ km s$^{-1}$). Furthermore, the $FWHM$ ($\Delta\varv$) is $\gtrsim$ 4 km s$^{-1}$ extending to 40 km s$^{-1}$. These ranges may be caused by inclination effects, which will be discussed further in Sect. \ref{sect:incl}. To quantify the emission, each profile is decomposed into Gaussian components, but leaving the deep absorptions seen in the ground-state lines masked out. The resulting offsets and $FWHM$ are shown in Fig. \ref{fig:vlsr_fwhm}. For each source, a characteristic offset velocity and $FWHM$ were chosen based on the decomposition of high-$S/N$ data without self-absorption, typically the 3$_{12}$--3$_{03}$ (1097 GHz) and 2$_{02}$--1$_{11}$ (988 GHz) transitions (Fig. \ref{fig:vlsr_fwhm}). These parameters were fixed and the decomposition redone for all spectra, letting only the intensity and the parameters of the secondary component be free. The secondary component is typically the broader component (except for the case of IRAS2A) associated with the outflow \citep{kristensen10, kristensen12}. The resulting intensities are listed in Table~\ref{tab:intensity}.

The main uncertainty in the listed intensities comes from the fitting and the uniqueness of the fit, particularly for low signal-to-noise lines. For strong components and very offset components, the fit is unique and the corresponding uncertainty low, as illustrated by the scatter of the width and offset shown in Fig. \ref{fig:vlsr_fwhm}. By fixing the width and offset we have removed this scatter by assuming that the shape of the component is independent of excitation. The decompositions obtained by fixing the width and offset are equally good to the decompositions where all parameters are free, where the quality of the fit is taken to be the residual. Typically, the rms of the residual is \mbox{$<$ 1.5} times the rms in a line-free region. 

\begin{figure}
\begin{center}
\includegraphics[width=0.7\columnwidth, angle=0]{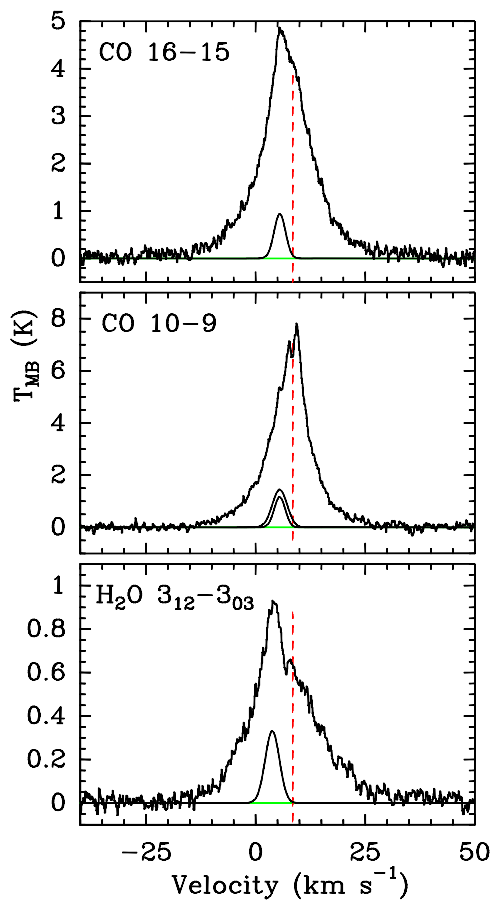}
\end{center}
\caption{Decomposition of the CO 10--9 profile towards SMM1 using either the best-fit parameters obtained from CO 16--15 (top) or H$_2$O (bottom). The Gaussian profiles illustrate the maximum amount of emission that can be hidden in the CO 10--9 profile.}
\label{fig:smm1_decomp}
\end{figure}

CO 10--9 spectra were also examined for the presence of an offset component by using the same kinematic parameters as in the H$_2$O decomposition. The strongest CO 10--9 line is observed towards SMM1 \citep{yildiz13} where no direct evidence is found for an offset component; the line profile can be decomposed without the need for an additional offset component. SMM1 does show a clear offset component in CO 16--15, and the question therefore arises, how much emission from the offset component can be hidden in the CO 10--9 profile? Figure \ref{fig:smm1_decomp} shows the CO 16--15, 10--9 and H$_2$O 3$_{12}$--3$_{03}$ spectra obtained towards SMM1. The offset component was first fitted using CO 16--15 and subsequently the $\Delta\varv$ and $\varv_{\rm LSR}$ were fixed and used to quantify emission in the CO 10--9 profile. Second, the same exercise was done but the offset parameters from the H$_2$O decomposition were fixed. There is little difference in terms of the integrated intensity no matter whether the CO or H$_2$O parameters are chosen (5.5 vs. 4.9 K km s$^{-1}$). The quality of the fits are the same as if the offset component is not included, and we therefore treat the inferred intensities as upper limits. Similarly for the other sources, only upper limits are available and these are based on the H$_2$O kinematic parameters (Table~\ref{tab:colimit}).

CH$^+$ and OH$^+$ spectra show the offset component in absorption (Benz et al. in prep.). OH$^+$ is detected towards all sources whereas CH$^+$ is detected towards two out of four sources. IRAS4B shows the CH$^+$ profile to be shifted both with respect to the source velocity and the offset component seen in H$_2$O; it is centred on 5.7--6.0 km s$^{-1}$ and is thus offset by more than 1 km s$^{-1}$ towards the blue. The CH$^+$ feature towards SMM1 is centred on the velocity of the offset component and has a larger $FWHM$ than the offset component seen in H$_2$O and CO (Fig.~\ref{fig:spectra}). C$^+$ is detected towards IRAS2A and SMM1, in both cases blue-shifted with respect to the source velocity by $\sim$ 4--5 km s$^{-1}$. 

OH is detected towards Ser SMM1 with HIFI at 1835 GHz (Wampfler et al. in prep.). A Gaussian decomposition is complicated by the fact that the hyperfine transitions are very closely spaced (2.4 km s$^{-1}$). Nevertheless, by fixing the intensity ratios of the hyperfine components, i.e., assuming the emission is optically thin and that the intensity ratios scale with $A_{\rm ul} g_{\rm u}$, a fit is obtained and the OH emission likely contains a mixture of the offset and broad component. Because of the shape of the profile, further decomposition is not performed, instead we assume that 50\% of the emission can be attributed to the offset component. HIFI OH spectra are not available towards the other sources as part of WISH.

\subsection{Time variability}

\begin{figure}
\begin{center}
\includegraphics[width=0.9\columnwidth, angle=0]{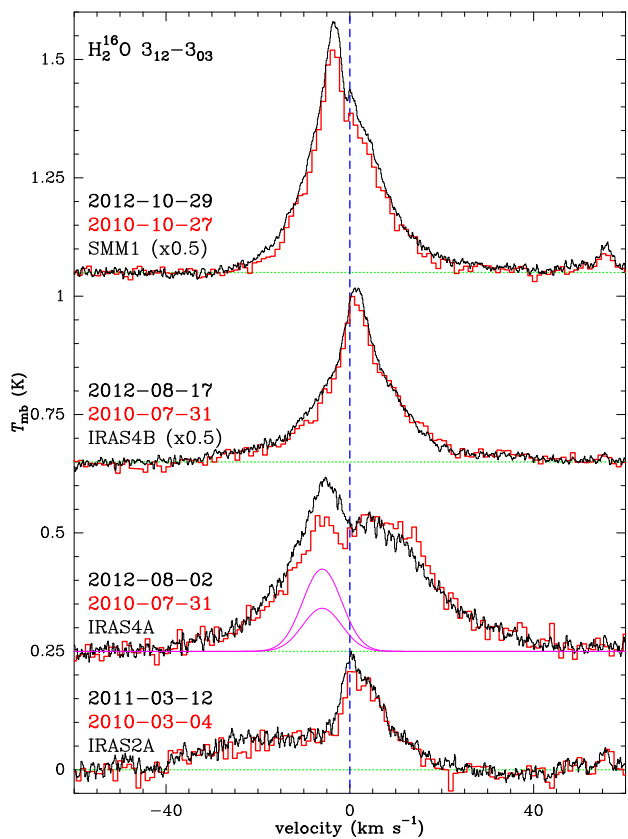}
\end{center}
\caption{H$_2$O 3$_{12}$--3$_{03}$ spectra at 1097 GHz towards SMM1, IRAS2A, IRAS4A and IRAS4B observed at two different epochs. The offset component towards IRAS4A has doubled in intensity, as shown by the two Gaussian fits in magenta. All spectra are shifted such that the source velocity is at 0 km s$^{-1}$.}
\label{fig:i4a_time}
\end{figure}

IRAS4A was re-observed in the H$_2$O 3$_{12}$--3$_{03}$ transition at 1097 GHz as part of an OT2 programme (PI: Visser) on Aug. 2, 2012, nearly two years after the original observations (July 31, 2010). During that period, the offset component doubled in intensity (Fig. \ref{fig:i4a_time}). IRAS2A, IRAS4B and SMM1 were also re-observed as part of the same OT2 programme, but show no signs of variability. All H$_2$O observations towards IRAS4A presented here were performed over a period of two months, and we assume that no significant variability took place over that time period. 

To verify that the change in emission is not caused by a pointing offset when the data were obtained, the pointing offsets were checked using HIPE 9.1. The recorded pointing offset towards IRAS4A was 3\farcs8 in July 2010 and less than 0\farcs5 in 2012. The pointing offsets were similar towards IRAS4B for both epochs and $\lesssim$ 2\farcs5 for both epochs for the other sources. For the pointing offset to be the cause of the intensity difference, the offset component would need to be located at the edge of the HIFI beam, i.e., at a distance of more than 10$''$ from the pointing centre to cause a doubling in intensity. Otherwise the pointing alone cannot account for the change. Below in Sect. \ref{sect:offset} we argue why this origin is unlikely, based on the hydride absorption. 

HIFI has an inherent calibration uncertainty of $\sim$ 10\% \citep{roelfsema12}. However, only the offset component shows a noticeable difference in intensity, the broader underlying outflow component appears unchanged between the two epochs. Towards SMM1, on the other hand, both the broad and offset components change intensity slightly, a change which can be attributed to calibration uncertainties; the difference in intensity is 10\% across the spectrum. In conclusion, the most likely explanation is that the offset component seen towards IRAS4A changed in intensity over the past two years. 

Spectra of species other than H$_2$O towards IRAS4A were obtained over a period of 6 months from March 3, 2010 to September 3, 2010, and we assume that little or no change took place in that time frame. 

\subsection{CO and limits on kinetic temperature}

\begin{figure}
\begin{center}
\includegraphics[width=\columnwidth, angle=0]{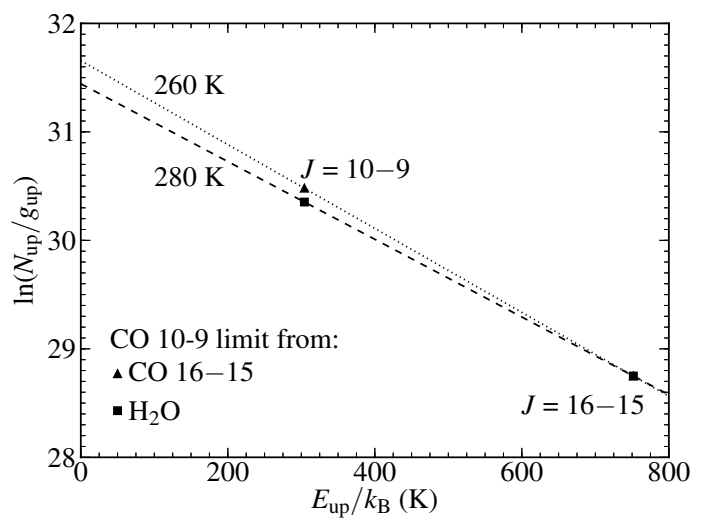}
\end{center}
\caption{CO rotational diagram for Ser SMM1 based on the upper limit of CO $J$ = 10--9 emission obtained from two different decompositions, and the detection in CO $J$ = 16--15. The lower limits on the rotational temperatures are shown for each upper limit on the CO $J$ = 10--9 emission.}
\label{fig:co_rotdiag}
\end{figure}

The offset component is seen in CO $J$=16--15 towards SMM1, but not in the lower-excited $J$=10--9 line. Thus, the upper limit from $J$=10--9 can be used to provide a lower limit on the rotational temperature and a corresponding upper limit on the column density, assuming the level populations are in local thermodynamic equilibrium and optically thin. The results are illustrated in Fig. \ref{fig:co_rotdiag}. The upper limit on the column density is $\sim$7$\times$10$^{13}$ cm$^{-2}$ in the 11\farcs5 beam of the CO 16--15 transition. 

Assuming LTE and that both lines are optically thin, the limit on the rotational temperature is $T_{\rm rot}$ $\gtrsim$ 270 K. \citet{gecco12} find that the CO ladder towards Ser SMM1 from $J$ = \mbox{4--3} to 49--48 consists of three rotational-temperature components with $T_{\rm rot}$ = 100, 350 and 600 K, respectively, corresponding to low-$J$, mid-$J$ and high-$J$ CO emission ($J$ $\lesssim$ 14, 26 and 42, respectively). The offset component is clearly not associated with the 100-K temperature component, but based on the limits on the rotational temperature it is not possible to conclude whether it is associated with the warm or hot component. Indeed, if the distribution is continuous  it is possible that the rotational temperatures do not correspond to discrete temperature regimes \citep[e.g.,][]{neufeld12}. 

For the specific case of SMM1 it is clear that the offset component is a distinct physical component based on the line profile, and this component must be present in the observed CO ladder. Moreover, the contribution is embedded in the integrated emission for $J_{\rm up}$ $>$ 10 which further illustrates the need for high spectral resolution observations to isolate emission from the separate dynamical components, as opposed to observations with, e.g., SPIRE and PACS on \textit{Herschel}. The same is likely the case for the other sources although a future analysis will show to what extent the CO $J$ = 10--9 data can be used to constrain the CO rotational temperature. If emission is strongly beam-diluted (see next section) it is likely that high angular resolution observations with a facility such as ALMA using CO $J$ = 6--5 will be able to further constrain the rotational temperature as well.

\begin{table}
\caption{CO 16--15 and limits on CO 10--9 integrated intensities in the offset components.}
\tiny
\begin{center}
\begin{tabular}{l c @{} c c c c}
\hline \hline
& CO 10--9 && \multicolumn{3}{c}{CO 16--15}  \vspace{1pt}  \\
\cline{2-2} \cline{4-6}
& $\int$ $T_{\rm MB}$ d$\varv$ && $\int$ $T_{\rm MB}$ d$\varv$ & $\varv_{\rm LSR}$ & $\Delta\varv$ \\
Source & (K km s$^{-1}$) && (K km s$^{-1}$) & (km s$^{-1}$) & (km s$^{-1}$) \\
\hline
IRAS2A & $<$ 0.9 && \ldots \\
IRAS4A & $<$ 1.2 && \ldots \\
IRAS4B & $<$ 6.6 && \ldots \\
SMM1(H$_2$O)\tablefootmark{a} & $<$ 4.9 && \ldots \\
SMM1(CO)\tablefootmark{b} & $<$ 5.5 && 6.2 & 5.5 & 3.4 \\
SMM3 & $<$ 2.3 && \ldots \\
\hline
\end{tabular}
\tablefoot{Upper limits are 1$\sigma$ and are obtained by fixing the position and width of the offset component from the H$_2$O data.
\tablefoottext{a}{$\varv_{\rm LSR}$ and $FWHM$ from the H$_2$O profiles.}
\tablefoottext{b}{$\varv_{\rm LSR}$ and $FWHM$ from the CO 16--15 profile.}}
\end{center}
\label{tab:colimit}
\end{table}

\subsection{H$_2$O and CO excitation conditions}\label{sec:excitation}

To determine the H$_2$O excitation conditions, $n$(H$_2$), $T$ and $N$(H$_2$O), specific H$_2$O line ratios are examined. The H$_2$O \mbox{3$_{12}$--3$_{03}$ / 3$_{12}$--2$_{21}$} ratio is particularly useful in providing initial constraints on the column density. Because the two transitions share the same upper level, the ratio is straightforward to calculate in the optically thin limit and is equal to 6.7. However, observations of these two transitions at 1153 and 1097 GHz, i.e., in similar beams, reveal an intensity ratio in the offset component ranging from 0.6 (IRAS4A) to 1.7 (SMM3). Such a ratio can only be explained if the 3$_{12}$--3$_{03}$ transition is optically thick ($\tau$ $>$ a few) and the column density is greater than $\sim$ 10$^{16}$ cm$^{-2}$ for any given emitting area.

\begin{figure*}
\sidecaption
\includegraphics[width=12cm, angle=0]{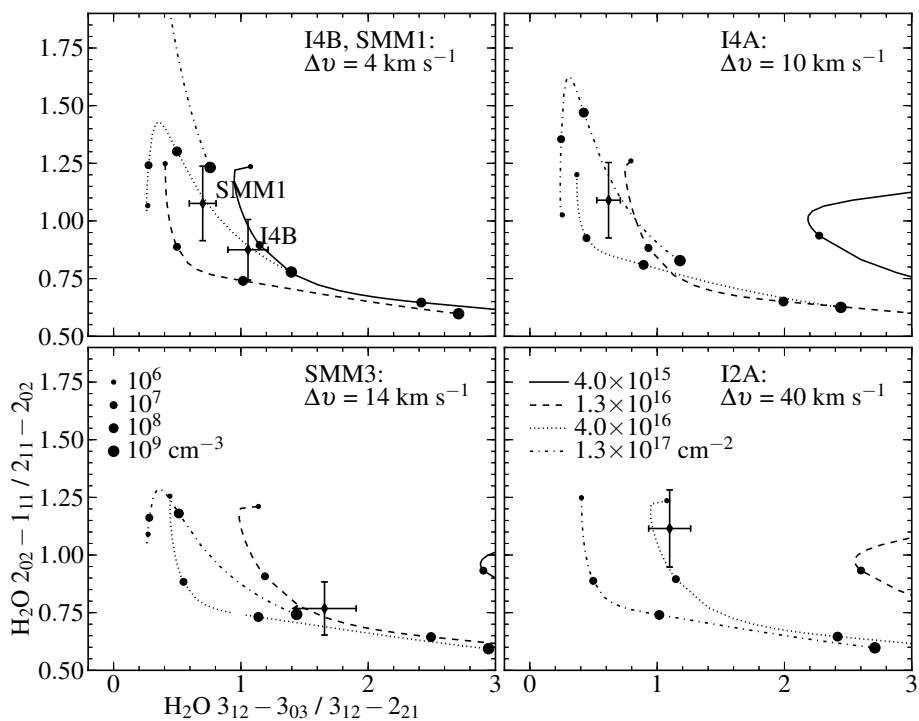}
\caption{H$_2$O 2$_{02}$--1$_{11}$ / 2$_{11}$--2$_{02}$ versus H$_2$O 3$_{12}$--3$_{03}$ / 3$_{12}$--2$_{21}$ line ratios for various H$_2$O column densities and H$_2$ densities from RADEX models. The four panels are for different line widths corresponding to the width of each of the offset components. The temperature is  fixed at 750~K. The different line styles correspond to different H$_2$O column densities and the dots are for different H$_2$ densities. The observed ratios are scaled to the same beam and marked with 15\% error bars in both ratios.}
\label{fig:h2o_diagnostic}
\end{figure*}

Figure \ref{fig:h2o_diagnostic} shows the H$_2$O 2$_{02}$--1$_{11}$ / 2$_{11}$--2$_{02}$ versus H$_2$O 3$_{12}$--3$_{03}$ / 3$_{12}$--2$_{21}$ line ratios for various H$_2$O column densities (4$\times$10$^{15}$--10$^{17}$ cm$^{-3}$), H$_2$ densities (10$^6$--10$^9$ cm$^{-3}$) and a temperature of 750 K. The ratios are calculated using the non-LTE statistical equilibrium code RADEX \citep{vandertak07} for line widths of $\Delta\varv$ = 4, 10, 14 or 40 km s$^{-1}$ corresponding to the $FWHM$ of the different offset components. The H$_2$O-H$_2$ collisional rate coefficients from \citet{daniel11} are used and the H$_2$ and H$_2$O o/p ratios are set to 3, the high-temperature equilibrium value. Observed line ratios are also shown and these have been scaled to the same 20$''$ beam assuming that the emitting region is much smaller than the beam. 

The resulting line ratios typically change by less than 10\% for temperatures in the range of 500--1000 K, i.e. they only weakly depend on the assumed temperature. \citet{gecco12} find from an excitation analysis of more H$_2$O lines that a kinetic temperature of $\sim$800 K reproduces the H$_2$O line ratios as well as the high-$J$ part of the CO ladder observed towards SMM1. We choose to fix the temperature to 750 K, the halfway point between 500 and 1000 K but note that this value is not constrained by the H$_2$O data.

For most model results there is a degeneracy between a (relatively) low H$_2$ density ($\sim$ 5 $\times$ 10$^6$ cm$^{-3}$), low H$_2$O column density (a few times 10$^{16}$ cm$^{-2}$) and a high H$_2$ density ($>$ 10$^8$ cm$^{-3}$), high H$_2$O column density ($>$ 10$^{17}$ cm$^{-2}$) (Fig. \ref{fig:h2o_diagnostic}). This degeneracy is most evident when comparing model results to the observations of IRAS4A, IRAS4B and SMM1, and corresponds to whether the line emission is sub-thermally or thermally excited. For the high column density case, the ground-state H$_2$O lines are very optically thick, $\tau$ $>$ 100, which may affect the accuracy of the radiative transfer. In the following, the results with the lowest column density and thereby lowest opacity will be analysed. The model results are summarised in Table \ref{tab:exc_con}.

The offset component was linked with 22 GHz H$_2$O maser emission in \citet{kristensen12}. For H$_2$O to mase, a density of $\sim$10$^7$ cm$^{-3}$ is required \citep{elitzur92}, which is typically a factor of two higher than what is inferred here. The maser density is based on H$_2$O-H$_2$ collisional rate coefficients which are more than twenty years old. Furthermore, the error bar on the observed line ratios is such that the best-fit densities span an order of magnitude and a density of 10$^7$ cm$^{-3}$ cannot be excluded. Thus the conclusion that the offset component is coincident with masers remains unchanged. 

The absolute H$_2$O 2$_{02}$--1$_{11}$ intensity from the RADEX models are compared to the observed intensity in the offset component to estimate the beam filling factor, or, alternatively, the radius of the emitting region, $r$. Finally, the CO column density is varied until the CO 16--15 intensity towards SMM1 is recovered for the same conditions as for H$_2$O. Typically, the radius of the emitting region is of the order of 100 AU (Table \ref{tab:exc_con}), or about $\sim$~0\farcs5 at a distance of 235 pc.

Towards SMM1, \citet{gecco12} obtain CO column densities of the warm ($T$ $\sim$ 375 K) and hot ($\sim$ 800 K) components of 10$^{18}$ cm$^{-2}$ and 5$\times$10$^{16}$ cm$^{-2}$, respectively, over a region with a radius of 500 AU. Note that these temperatures are inferred from modelling the CO ladder and H$_2$O emission, and are not identical to the measured CO rotational temperature \citep[600~K,][]{gecco12}. If our inferred CO column density of 10$^{18}$ cm$^{-2}$ over a 110 AU emitting radius is scaled to a radius of 500 AU, the CO column density becomes $\sim$ 5$\times$10$^{16}$ cm$^{-2}$. It is therefore likely that the hot component observed in the high-$J$ CO data is identical to the offset component identified in the HIFI H$_2$O and CO data; the column density of the warm CO component is too high to be hidden in the HIFI offset component. This analysis shows that the profiles are necessary for disentangling the different kinematical components in each spectrum, and that the hot CO detected with PACS is likely a distinct physical component towards this source. 

This work assumes that the temperature of the H$_2$O emitting gas is the same as that of CO, and that the temperature of the H$_2$O emitting gas is close to what has been determined by, e.g., \citet{gecco12}. What if this is not the case? Furthermore, how much column density can be hidden in the CO 10--9 profiles, where the offset component is not detected? The CO $J$=10--9 contribution of the warm and hot components is estimated from the rotational diagrams in \citet{karska13}, \citet{herczeg12} and \citet{gecco12}. In all cases, the integrated CO $J$=10--9 intensity is of the order of $\sim$30--40 K km s$^{-1}$ for the warm component and $\sim$ 2.5--3 K km s$^{-1}$ for the hot component when extrapolating the linear fits from \mbox{$J$ $\sim$ 15--40} down to $J$ = 10 and assuming emission is optically thin. The opacity can be estimated from optically thin $^{13}$CO 10--9 emission and the $J$ = 10--9 line is optically thin away from the line centre \citep{sanjosegarcia13, yildiz13}. The upper limit for the offset component in the CO 10--9 data is \mbox{$\sim$ 1--6 K km s$^{-1}$} (Table \ref{tab:colimit}). Thus, the offset component would have easily been detected in the HIFI spectra if it were associated with the warm PACS component, but not if it is associated with the hot component. 

The H$_2$O/CO abundance ratio towards SMM1 is $\sim$0.04. Our analysis assumes that CO and H$_2$O share excitation conditions. This may not be the case, as CO is shifted with respect to H$_2$O by $\sim$ 1.5 km s$^{-1}$ towards SMM1, although they have identical line widths. \citet{gecco12} find a higher H$_2$O/CO abundance ratio of 0.4 for the same H$_2$ density of 5$\times$10$^6$ cm$^{-3}$, but for a different emitting region. Since little H$_2$ data exist towards the central source position, and certainly no H$_2$ data with the velocity resolution required to isolate the offset component, the H$_2$O/CO ratio serves as a proxy for the H$_2$O abundance with respect to H$_2$. For a canonical CO abundance of 10$^{-4}$ the H$_2$O abundance is 4$\times$10$^{-6}$ and thus lower by several orders of magnitude compared to what would be expected if all oxygen were locked up in H$_2$O ($\sim$ 3$\times$10$^{-4}$).

\begin{table}
\caption{H$_2$O and CO excitation conditions.}
\begin{center}
\begin{tabular}{l c c c c c}
\hline \hline
Source & $n$(H$_2$) & $N$(H$_2$O)\tablefootmark{a} & $N$(CO)\tablefootmark{a} & $T_{\rm kin}$\tablefootmark{b} & $r$ \\
& (cm$^{-3}$) & (cm$^{-2}$) & (cm$^{-2}$) & (K) & (AU) \\
\hline
IRAS2A	& 5$\times$10$^6$ & 4$\times$10$^{16}$ &  & 750 & 80 \\
IRAS4A	& 5$\times$10$^6$ & 1$\times$10$^{16}$ &  & 750 & 140 \\
IRAS4B	& 1$\times$10$^7$ & 4$\times$10$^{15}$ &  & 750 & 160 \\
SMM1	& 5$\times$10$^6$ & 4$\times$10$^{16}$ & 1$\times$10$^{18}$ & 750 & 110 \\
SMM3	& 5$\times$10$^7$ & 1$\times$10$^{16}$ &  & 750 & \phantom{1}50 \\
\hline
\end{tabular}
\tablefoot{
	\tablefoottext{a}{Column density over the emitting region with radius $r$.}
	\tablefoottext{b}{Kinetic temperature fixed in the model.}
}
\end{center}
\label{tab:exc_con}
\end{table}

\subsection{OH$^+$, C$^+$ and CH$^+$}

Observations and characteristics of the hydride observations are reported in Benz et al. (in prep.). We here summarise the main results concerning the offset component and report again the hydride column densities for completeness.

The offset component is uniquely identified in absorption in both OH$^+$ and CH$^+$ towards IRAS4B and SMM1, and in OH$^+$ towards all sources. The velocity offsets are consistent with those seen in H$_2$O and, for the case of SMM1, in CO $J$ = 16--15. From the absorption features it is possible to directly measure the absorbing column through 
\begin{equation}
N_{\rm low} = \frac{8\pi}{c^3} \frac{\nu^3 g_{\rm low}}{A_{\rm ul} g_{\rm up}} \int \tau\, {\rm d}\varv\ ,
\end{equation}
where $\nu$ is the line frequency, $A_{\rm ul}$ the Einstein $A$-coefficient and $g$ the statistical weight of the lower and upper levels. The opacity, $\tau$, is determined as $\tau$ = ln($T_{\rm cont}$ / $T_{\rm line}$). In determining the column density, it is implicitly assumed there is no re-emission of the absorbed photons. The measured values are given in Table \ref{tab:hydrides} along with 3$\sigma$ upper limits (Benz et al. in prep.).

The CH$^+$ column densities are in the range of $<$ a few times 10$^{11}$ cm$^{-2}$ to 2 $\times$ 10$^{13}$ cm$^{-2}$. This range is similar to that found in diffuse interstellar clouds \citep[$\sim$ 10$^{12}$--10$^{14}$ cm$^{-2}$;][]{gredel97}. The OH$^+$ column densities are of the order of 10$^{13}$ cm$^{-2}$. The OH$^+$ column density is similar to what \citet{bruderer10} measured towards the high-mass star-forming region AFGL2591, $N$(OH$^+$) $\sim$ 1.6 $\times$ 10$^{13}$ cm$^{-2}$, whereas the CH$^+$ column densities towards the low-mass objects are 1--2 orders of magnitude lower than towards AFGL2591, $N$(CH$^+$) $\sim$ 1.8 $\times$ 10$^{14}$ cm$^{-2}$.

C$^+$ is not uniquely identified with the offset component although an absorption feature is seen towards SMM1 at the velocity of the offset component, and towards IRAS2A closer to the source velocity. The measured column densities are 4$\times$10$^{17}$~cm$^{-2}$. The observations were performed in dual-beam-switch mode and it cannot be ruled out that some emission is missing because it is chopped out. For example, \citet{larsson02} observed extended [\ion{C}{ii}] emission over the entire Serpens core with ISO-LWS. However, the absorption is expected to mainly affect any kinematical component at or close to the source velocity; the offset component is blue-shifted by several km s$^{-1}$ which suggests that the [\ion{C}{ii}] absorption is unrelated to the cloud or Galactic foreground but intrinsic to the sources.

\begin{table}
\caption{Column densities of OH$^+$, C$^+$ and CH$_+$ where detected, and 3$\sigma$ upper limits on CH and HCO$^+$ column densities. }
\begin{center}
\scriptsize
\begin{tabular}{l c c c c c}
\hline \hline
Source & $N$(OH$^+$) & $N$(CH$^+$) & $N$(C$^+$) & $N$(HCO$^+$) & $N$(CH) \\
& (cm$^{-2}$) & (cm$^{-2}$) & (cm$^{-2}$) & (cm$^{-2}$) & (cm$^{-2}$)  \\
\hline
IRAS2A	& 2.2$\times$10$^{13}$ & $<$4.2$\times$10$^{11}$ & 3.9$\times$10$^{17}$ & $<$8.0$\times$10$^{13}$ & $<$3.7$\times$10$^{14}$ \\
IRAS4A	& 1.5$\times$10$^{13}$ & $<$6.4$\times$10$^{11}$ & $<$9.6$\times$10$^{16}$ & $<$6.6$\times$10$^{13}$ & $<$3.1$\times$10$^{14}$ \\
IRAS4B	& 7.5$\times$10$^{12}$ & 1.8$\times$10$^{12}$ & $<$3.9$\times$10$^{17}$ & $<$2.0$\times$10$^{14}$ & $<$9.4$\times$10$^{14}$ \\
SMM1	& 3.3$\times$10$^{13}$ & 2.4$\times$10$^{13}$ & 4.2$\times$10$^{16}$ & $<$3.3$\times$10$^{14}$ & $<$1.5$\times$10$^{15}$ \\
SMM3	& \ldots & \ldots & \ldots & $<$1.1$\times$10$^{15}$ & $<$4.9$\times$10$^{15}$ \\
\hline
\end{tabular}
\tablefoot{OH$^+$ and CH$^+$ are from Benz et al. (in prep.). SMM3 was not targeted for observations of OH$^+$, CH$^+$ and C$^+$.
}
\end{center}
\label{tab:hydrides}
\end{table}

\subsection{OH, CH and HCO$^+$}

The offset component is not uniquely identified in the spectra of OH, CH and the very deep HCO$^+$ $J$ = 6--5 line obtained serendipitously in our observations of H$_2^{18}$O 1$_{10}$--1$_{01}$ \citep{kristensen10}. Nevertheless, for the OH HIFI spectrum towards SMM1, we assume that 50\% of the emission and therefore 50\% of the column density can be assigned to the offset component (see discussion above, Sect. \ref{sect:gauss}). An OH column density of 5$\times$10$^{15}$ cm$^{-2}$ is adopted, a value which is probably accurate to within a factor of a few \citep{wampfler13}.

As for CH and HCO$^+$, emitting region sizes, temperatures, H$_2$ densities and line-widths are as for H$_2$O and CO. For the case of CH, only one transition is observed and we therefore adopt the upper limit on the rotational temperature from \citet{bruderer10} of 25 K to estimate the total column density. If CH is in LTE and is optically thin with a rotational temperature of $>$ 500 K, the upper limit is only a factor of $\sim$ 5 higher than what is given in Table \ref{tab:hydrides}. The same procedure is adopted for HCO$^+$ 6--5 with the exception that the rotational temperature is taken to be $>$ 500 K, because the critical density of HCO$^+$ is $\sim$2$\times$10$^7$ cm$^{-3}$ and so emission may be in LTE. If HCO$^+$ is strongly subthermally excited and the rotational temperature is only 25 K, the column densities are overestimated by $\sim$ 25\%. Typical upper limits are $\sim$ 10$^{14}$ cm$^{-2}$ and 10$^{15}$ cm$^{-2}$ for HCO$^+$ and CH, respectively.

\section{Discussion}
\label{sec:disc}

\subsection{Location of the offset component}\label{sect:offset}

Because the offset component is typically blue-shifted and because it sometimes appears in absorption in certain species against the continuum, it is possible to constrain the physical location of the component in the protostellar system. First, we assume that the offset component consists of both a red- and blue-shifted component, but that the red-shifted component is hidden from view by some obscuring agent. This obscuring agent may consist of either gas or dust (or both), and both possibilities are discussed below.

The CO 16--15 emission is optically thin \citep{gecco12} and the red-shifted counterpart is the most difficult to hide. Is it possible to have a layer of CO gas between the blue- and red-shifted offset components that could shield the red-shifted component, and if so, what would the conditions need to be? For high densities of 5$\times$10$^6$ cm$^{-3}$, the optical depth of the CO 16--15 line is nearly independent of density, and thus only depends on temperature and column density. For temperatures greater than $\sim$ 100 K, the optical depth scales almost linearly with temperature. Thus, for $N$(CO) = 3$\times$10$^{18}$ cm$^{-2}$ and $T$ = 750 K the CO 16--15 transition is optically thick with $\tau$ = 3, for $\Delta\varv$ = 4 km s$^{-1}$, the width appropriate for SMM1. However, such conditions yield significant emission in the lower-$J$ transitions, and although the component would remain hidden in higher-$J$ lines, it would appear in lower-$J$ lines. Only for low temperatures and high column densities, e.g., $N$(CO) = 3$\times$10$^{19}$ cm$^{-2}$ and $T$ = 100 K is emission in both the high-$J$ and low-$J$ lines optically thick, similar to the conditions observed in a Herbig Ae/Be disk \citep{bruderer12} where even CO 16-15 is marginally optically thick. For H$_2$ densities of 5$\times$10$^6$ cm$^{-3}$, a column length of $\gtrsim$ 400 AU is required to obscure this component from view. If the density of the gas is higher, 10$^8$ cm$^{-3}$, the column length is correspondingly lower, 100 AU. Such length scales correspond to the inferred sizes of embedded disks around Class 0 objects \citep[e.g.,][]{jorgensen07, jorgensen09} and is also similar to the inferred size of the emitting region (Table \ref{tab:exc_con}). Gas densities in excess of 10$^8$ cm$^{-3}$ are only found close to the protostar in the disk, where the temperature is low. Thus it is possible that the red-shifted counterpart is hidden by large amounts of high-density, cold CO gas.

The red counterpart of the offset component is also obscured in H$_2$O emission. However, for similar conditions as discussed above ($n$ = 5$\times$10$^6$ cm$^{-3}$, $T$ = 100 K) an H$_2$O column of only $\sim$ 10$^{16}$ cm$^{-2}$ is required to obscure material. If the gas and dust temperatures are equal and close to 100 K, the water abundance is expected to be within a factor of a few of the CO abundance, and thus it is straightforward to hide any water emission appearing on the red side of the spectrum by colder H$_2$O gas.

The shielding molecular gas can be associated with the entrained gas in the molecular outflow. If that is the case, then a significant fraction of the outflow is entrained on very small scales ($<$ a few hundred AU) in order to hide the red offset component which also resides in the inner few hundred AU. Alternatively, the obscuring gas originates in the infalling envelope. Free-falling gas towards a protostar with a mass of 0.5 $M_\odot$ has a velocity of 3 km s$^{-1}$ at a distance of 100 AU, and therefore it is possible that the gas flowing towards the protostar (red-shifted and located on the side facing us) shields the red offset component located on the far side of the system in the sources where the offset and width is low. For the case of IRAS2A where the offset is 13 km s$^{-1}$ and the width is 40 km s$^{-1}$, the infalling gas cannot shield the red-shifted offset component.

\begin{figure}
\begin{center}
\includegraphics[width=\columnwidth, angle=0]{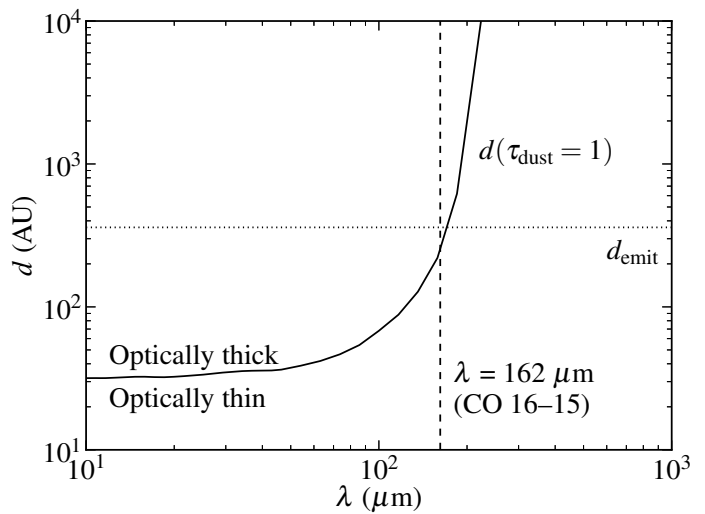}
\end{center}
\caption{Thickness of the optically thick dust zone as a function of wavelength from the inside of the envelope of SMM1 towards the outside using the model envelope of \citet{kristensen12}. The diameter of the emitting region is marked. }
\label{fig:tau1}
\end{figure}

Shielding by the dust is another possibility. The lowest frequency at which the offset component is detected is at 557 GHz in the H$_2$O 1$_{10}$--1$_{01}$ transition. Adopting a dust opacity of 5 cm$^2$ g$^{-1}$, the opacity from Table 1, column 5 of \citet{ossenkopf94} at 500 $\mu$m, a gas/dust ratio of 100 and a mean molecular weight of 2.8 $m_{\rm H}$, an H$_2$ column density of $>$ 10$^{24}$ cm$^{-2}$ is required for a dust optical depth of 1. At shorter wavelengths, the dust opacities increase and thus a lower dust column density is required to shield the offset component. Figure \ref{fig:tau1} illustrates the dust $\tau$ = 1 surface as a function of wavelength from the inside of the envelope of SMM1 using the spherical envelope model of \citet{kristensen12}. In this representation an observer is able to see the other side of the envelope if he is located in the ``optically thin'' zone; in the ``optically thick'' zone the dust blocks emission from the other side. At 162 $\mu$m, the wavelength of the CO $J$=16--15 transition, an H$_2$ column density of more than 10$^{23}$~cm$^{-2}$ is required for the dust to be optically thick, which is obtained on scales of $\sim$ 220 AU in this model, i.e., comparable to the size of the emitting region (Fig. \ref{fig:tau1}). Therefore it is possible that dust obscures the red-shifted offset component at 162 $\mu$m, but this is not the case at 500 $\mu$m.

 \citet{jorgensen05a} showed that on scales of a few hundred AU, an additional component is required to reproduce interferometric observations. This component is likely the protostellar disk, and the column density is sufficient to provide enough shielding from the far side \citep[see][for the example of SMM1]{enoch09}. In conclusion, the red-shifted component is likely hidden either by high-density molecular gas in the disk or the inner, dense envelope. 

The fact that several species tracing the offset component (e.g., OH$^+$, CH$^+$ and H$_2$O towards IRAS3A and SMM3) only appear in absorption against the continuum from the disk/envelope places the offset component in front of the disk. The continuum-emitting region at the wavelengths where these absorptions appear (300 $\mu$m for OH$^+$ to 540 $\mu$m for H$_2$O) is typically up to $\sim$~500~AU in size \citep[$\sim$ 2\arcsec;][]{jorgensen09, gecco12}. These considerations all point to a physical origin of the offset to within the inner few hundred AU of the protostar, and located between us and the protostar itself.

\subsection{Inclination}\label{sect:incl}

The three NGC1333 sources have well-constrained inclination angles with respect to the plane of the sky. The outflow from IRAS2A is close to the plane ($i$ $\sim$ 90\degr), IRAS4A is at $i$ $\sim$ 45\degr, and IRAS4B is seen nearly pole on ($i$ $\sim$ 0\degr). Fitting the spectral energy distribution of SMM1, \citet{enoch09} find that SMM1 has an inclination of 30\degr. IRAS2A has the largest offset and $FWHM$ whereas that towards IRAS4B shows the smallest offset and $FWHM$, with IRAS4A and SMM1 falling between these two extremes. This four-point correlation suggests that the offset and width depend on inclination, and that the offset component is moving nearly perpendicular to the large-scale outflow as explained below.

If the origin of the offset component is a shock as suggested by the width and offset of the profile, how will the shape of the profile depend on the inclination? The offset velocity follows the inclination as sin($i$); when the inclination is 0\degr\ (the case of IRAS4B) the offset is 0 km s$^{-1}$ whereas it reaches its maximum value at $i$ = 90\degr\ (IRAS2A). The width will be narrow when observing the shock from an orientation close to face-on (IRAS4B) because only the velocity component inherent to the shock is probed. For an edge-on orientation the profile will appear broader because the shock is now observed at inclinations ranging from the plane of the sky to the line of sight. 

\begin{figure}
\begin{center}
\includegraphics[width=\columnwidth, angle=0]{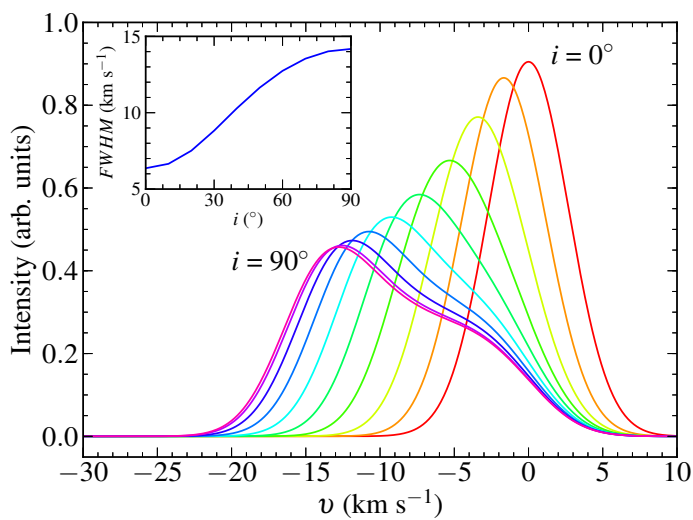}
\end{center}
\caption{Toy model of line profiles originating in an expanding half annulus as a function of inclination to the observer. An inclination of 0\degr\ corresponds to the annulus expanding into the plane of the sky (IRAS4B) and 90\degr\ corresponds to expansion along the line of sight (IRAS2A). The incremental inclination is 10$^\circ$. The inset shows the measured $FWHM$ of the different profiles as a function of inclination, i.e., not obtained from a Gauss fit. The annulus expands at 15 km s$^{-1}$ and the internal velocity dispersion ($FWHM$) is 9 km s$^{-1}$.}
\label{fig:vel_inc}
\end{figure}

To illustrate how the velocity offset and width changes with inclination in the proposed scenario, we make a geometrical toy model. The model consists of a half annulus which expands from the plane of the sky towards the observer. The intensity from any given point along the annulus corresponds to a Gaussian with a predefined offset (expansion velocity) and width (internal velocity dispersion) which both stay constant along the annulus. The offset, however, moves and an expansion into the plane of the sky corresponds to an offset velocity of 0 km s$^{-1}$, i.e., the offset velocity along the annulus scales with the angle $\theta$ going from 90\degr\ (the line of sight) to 0 and 180\degr\ (the plane of the sky). The annulus is furthermore given an inclination to the plane of the sky, $i$, which ranges from 0\degr\ (the plane of the sky) to 90\degr\ (the line of sight). The resulting profiles from the expanding half annuli are shown in Fig. \ref{fig:vel_inc}, where also the evolution of the width with inclination is shown. No single set of parameters (offset and width) are able to reproduce all observations, either because no such single set exists, or the toy model is too simplistic to capture the geometry and excitation. For example, here only a single shock or expansion velocity is considered; a range of velocities may be more appropriate as in the case of bow shocks \citep[e.g.][]{kristensen07}. Nevertheless, the behaviour of the offset components is captured qualitatively which suggests that the wind scenario is a geometrically possible solution, and that a shock velocity of 15~km~s$^{-1}$ is reasonable from the line profile perspective.

\subsection{Physical origin}\label{sec:origin}

Both the offset (2--15 km s$^{-1}$) and the width ($\sim$ 4--40 km s$^{-1}$) are indicative of a shock origin, a shock appearing close to the protostar. \citet{neufeld89} modelled fast, dissociative shocks and included the effects of UV radiation generated in the shock itself through Ly$\alpha$ emission. After initial heating to $>$ 5$\times$10$^4$ K, the compressed, shocked gas cools to $T$ $\sim$ 5000~K where it reaches a plateau. During this phase, the electron abundance is high ($\gtrsim$10$^{-2}$) and molecular ions are abundant, e.g., OH$^+$ and CH$^+$.  The gas is compressed by a factor of $\sim$ 400 in this stage to $\sim$ 10$^8$ cm$^{-3}$. Eventually the OH formation rate exceeds the destruction rate, and OH brings the temperature down to $\sim$~500 K, at which point the temperature reaches another plateau while H$_2$ forms. Once H$_2$ is formed, the temperature quickly drops to $\sim$ 100 K when CO and H$_2$O take over as dominant coolants. 

\begin{figure}
\begin{center}
\includegraphics[width=\columnwidth, angle=0]{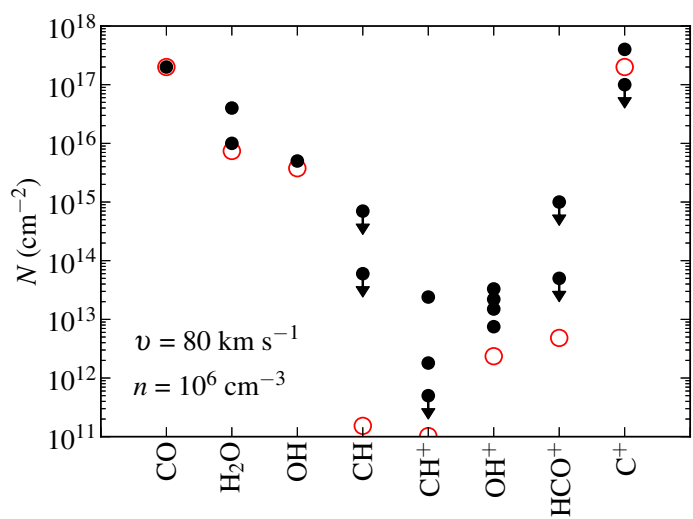}
\end{center}
\caption{Comparison of inferred column densities over the size of the emitting region and upper limits with shock model results from \citet{neufeld89} for a pre-shock density of 10$^6$ cm$^{-3}$ and shock velocity of 80 km s$^{-1}$. Observations are marked with black dots and arrows are for upper limits. For the cases where both a black dot and arrow are present (OH$^+$ and C$^+$) the dot marks the detection and the arrow the upper limit towards the other sources; when two arrows are present (CH and HCO$^+$) they illustrate the range of upper limits. Model results are shown as red circles and are normalised to the inferred CO column density.}
\label{fig:nd89_comp}
\end{figure}

The predicted column densities \citep{neufeld89} are in good agreement with the inferred observational column densities (Fig. \ref{fig:nd89_comp}) for a dense, dissociative shock (10$^6$ cm$^{-3}$) with a velocity of 80 km s$^{-1}$. There is a trend in the model predictions for higher column densities of C$^+$ with higher velocity, but column densities are not reported for $\varv$ $>$ 60 km s$^{-1}$ and \mbox{$n$ = 10$^6$ cm$^{-3}$} (C$^+$ and CH column densities are taken from the model with $\varv$ = 60 km s$^{-1}$). In general, the agreement is remarkable and shows that a fast, dense shock is a possible explanation for the observed column densities. 

The high density required by the model is easily found in the inner parts of the molecular envelope. The high velocity required is probably attained in either the jet or the strong wind from the protostar, although no direct observations exist of the wind in Class 0 objects. In the following, the ``jet'' refers to the highly collimated and fast component observed as extremely high-velocity features in molecular species, and the ``wind'' refers to the wide-angle, slower component seen towards Class I and II sources \citep[e.g.,][and references therein]{arce07}. The slower wind is primarily observed in forbidden atomic and ionic transitions at near-infrared and shorter wavelengths \citep{arce07, ray07} where the velocity is $\lesssim$ 10--20 km s$^{-1}$. 

\begin{figure}
\begin{center}
\includegraphics[width=\columnwidth, angle=0]{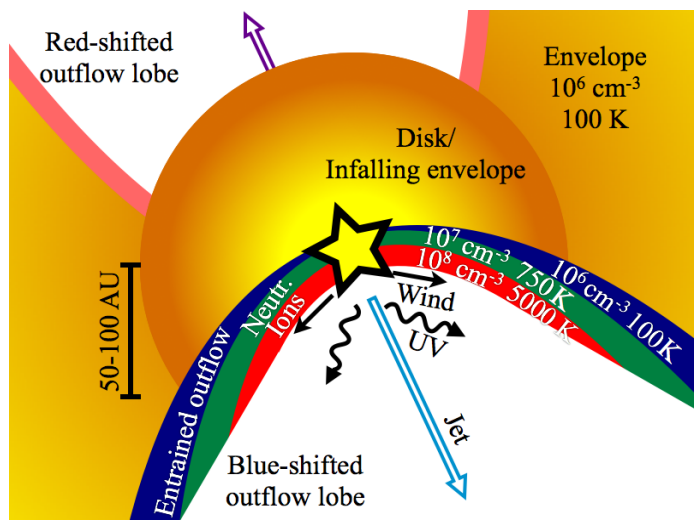}
\end{center}
\caption{Cartoon illustrating the inner few 100 AU of a low-mass protostar (not to scale). Where the wind and UV photons interact directly with the envelope, an irradiated shock occurs which compresses the envelope by several orders of magnitude over short distances. On the side of the cavity walls facing the protostar, ions (C$^+$, CH$^+$, OH$^+$) dominate emission and cooling. Neutral species (OH, H$_2$O, CO) dominate further into the envelope and on larger size scales. The disk hides the innermost parts of the red-shifted outflow lobe. Typical densities and temperatures are provided where relevant.}
\label{fig:cartoon}
\end{figure}

If the shock is moving at 80 km s$^{-1}$ perpendicular to the outflow direction (see above, Sect. \ref{sect:incl}) the envelope will quickly dissipate on timescales of $<$ 10$^3$ years \citep[e.g.,][]{shang06} and the wind would be much faster than what is observed at later evolutionary stages. The key ingredients that a successful model should reproduce are: \textit{(i)} the excitation conditions and \textit{(ii)} the hydride column densities, while \textit{(iii)} not dissipating the envelope on very short timescales. The reason the fast dissociative shock is successful in reproducing the first two ingredients is the layered structure where the hottest gas is atomic (dissociated) and as the gas cools, molecules reform. Two alternatives, which would also reproduce the third ingredient, are possible: \textit{(a)} either the dissociating UV photons are not intrinsic to the shock in which case the shock velocity can be lower, possibly as low as the wind velocity in Class I/II sources (10--20 km s$^{-1}$), see Fig. \ref{fig:cartoon} for a schematic of the scenario; or \textit{(b)} the wind is faster in Class 0 sources but only interacts with the envelope in very small regions and over short enough timescales that no irreparable damage is done to the large-scale envelope, i.e., the wind is anisotropic in time and direction. In the following we argue why a combination of the two is a likely solution.

Accreting protostars generate copious amounts of UV photons \citep[e.g.,][]{ingleby11} and thus provide a natural source of external UV illumination \citep{spaans95, vankempen09a, visser12, yildiz12}. If the UV field is strong and hard enough, the ion chemistry naturally evolves in a similar fashion to high-mass star-forming objects along the outflow cavity walls \citep[e.g.,][]{stauber07, bruderer09, bruderer10, benz10}, and the high shock velocity is not required to dissociate the molecules. \citet{visser12} argue that the cavity density is low enough and that the distance to the cavity wall from the protostar is short enough that UV photons reach the cavity wall mostly unattenuated. Depending on the actual shock conditions and the characteristics of the UV field, it is possible that the combination of a dissociating UV field and lower shock velocity similar to the wind velocities observed towards Class I and II sources ($\varv$ $\sim$ 10--20 km$^{-1}$) can reproduce the chemical and physical conditions. However, models of dense irradiated shocks are required to test this hypothesis further.

The variability observed towards IRAS4A demonstrates that heating and cooling in the offset component takes place on timescales of years. Such a variability is only possible in J-type shocks \citep{flower03}, where the cooling lengths are short enough, of the order of a few tens of AU or less. This is consistent with the dynamical distance, i.e., the distance a shock with a velocity of 10 km s$^{-1}$ traverses over 2 years ($\sim$~5~AU). Furthermore, the variability illustrates that the offset components are transient phenomena, and that the driving agent is not constant over time, i.e., the envelope may have time to settle between outburst events. The combination of a low shock velocity \mbox{(10--15 km s$^{-1}$)}, dense medium (10$^6$--10$^8$ cm$^{-3}$), UV-driven chemistry and time variability point to a scenario in which shocks from the protostellar wind impinge on the very inner dense envelope at angles that are close to perpendicular to the large-scale outflow. Whether the envelope has enough time to relax between outburst events or not will await further observations.

\subsection{Tracing winds and shocks in other sources}

Most of the neutral species are reformed at a temperature plateau in the shock of $\sim$ 500 K. The heating of this plateau region is dominated by H$_2$ formation on grains, and CO, H$_2$O, OH and other neutral species dominate the cooling. The temperature of this plateau is coincident with that of the hot component seen in the PACS high-$J$ CO data towards low-mass protostars \citep{herczeg12, gecco12,manoj12,karska13, green13}, and it is possible that the CO emission arises in the dense post-shock gas. This would imply that almost all protostars contain dense dissociative shocks, and that the ``universality'' of this temperature component is due to the energy release of H$_2$ reformation. 

If this interpretation is correct, it provides geometrical constraints on the type of system where the offset component can be observed. First, the offset component needs to move out of the plane of the sky, or the radial velocity is zero. If the component is moving entirely in the plane of the sky, the component is incident with the source velocity and there is no or little offset as in the case of IRAS4B. Second, the infall rate needs to be large enough that the red offset component is shielded by the infalling gas or a more dense inner envelope or disk. If it is not shielded, the profile becomes symmetric around the source velocity making identification more difficult. The offset component is only detected towards the more massive envelopes in the WISH low-mass sample with the higher infall rates and larger disks, which corroborates this interpretation. 

If the infall rate is very low, the corresponding outflow rate is low and therefore the offset component will be weak because the shock is weaker. The sources showing the offset component have the highest mass of hot gas as measured by the very high-$J$ CO emission observed with PACS \citep{herczeg12,karska13}, suggesting that the other reason the offset component is not observed towards more sources may be a question of the signal-to-noise ratio. Alternatively, since the variability observed towards IRAS4A takes place over timescales of a few years, the lack of an offset component indicates that the current shock activity is low.

The offset component is primarily seen in H$_2$O, and as illustrated above, H$_2$O is a better tracer of the kinematics than CO when it comes to hot shocked gas. However, two additional reasons to those mentioned above may play a role as to why an offset component is not seen towards all sources in H$_2$O. First, when the envelope is diluted UV photons may penetrate further into the envelope and photodissociate H$_2$O even in the post-shock gas, thereby lowering the H$_2$O column density \citep{nisini02, visser12, karska13}. Second, as the H$_2$ density becomes lower excitation of H$_2$O becomes more inefficient, and thus the $S/N$ decreases significantly \citep{nisini02, kristensen12}. Thus, the reason the offset component is not detected towards all sources is likely a combination of geometrical effects and low signal from the emitting gas.

\section{Summary and conclusions}

\begin{itemize}
\item Water emission profiles trace components not seen in low-$J$ CO ($J$ $<$ 10--9) observed from the ground and space-based telescopes such as \textit{Herschel}. These components are uniquely detected in ionised hydrides such as OH$^+$ and CH$^+$, but also OH, CO 16--15 and C$^+$ show the same component in absorption and emission. The component is not detected in deep spectra of HCO$^+$ 6--5 nor CH, thus providing valuable upper limits on the column densities. 
\item The component is observed at two epochs towards most sources and the intensity doubled over a period of two years in the H$_2$O 3$_{12}$--3$_{03}$ transition at 1097 GHz towards IRAS4A. The variability points to an origin in gas which is rapidly heated or cooling.
\item H$_2$O column densities are estimated to be $\gtrsim$ 10$^{16}$ cm$^{-2}$ while the H$_2$ density is 5$\times$10$^6$ -- 5$\times$10$^7$ cm$^{-3}$. The emitting regions are small, $\sim$ 100 AU. From the detection of the offset component in CO 16--15 the CO column density over the emitting region is estimated to be 10$^{18}$ cm$^{-2}$. The C$^+$, CH$^+$ and OH$^+$ column densities are measured and upper limits on CH and HCO$^+$ are provided.
\item The inferred CO column density implies that the offset component is identical to the hot CO component \mbox{($T$ $\sim$ 700--800~K)} seen in \textit{Herschel}-PACS data towards these low-mass protostars. The temperature is caused by the reformation of H$_2$ in the post-shock gas.
\item Several of the light hydrides are observed in absorption against the continuum, directly pointing to an origin close to the protostar. The most likely origin of this component is in dissociative shocks where the dissociation may come from the UV light from the accreting protostar rather than the shock itself, thus lowering the required shock velocity. The shocks possibly arise in the interaction between the protostellar wind and the inner, dense envelope. The UV field is required to account for the large column densities of the ionised hydrides, and models of irradiated shocks are necessary to further test and quantify this hypothesis. 
\end{itemize}

These observations highlight the unique capability of, in particular, H$_2$O as a tracer of dynamical components in protostars even on the smallest spatial scales. Specifically, the data unequivocally reveal that dissociative shocks in the inner $\sim$~100~AU of low-mass protostars are common. These observations thus open a window towards understanding how low-mass protostars energetically interact with their parental material. Future facilities such as ALMA will undoubtedly shed more light on these energetic processes for a larger number of sources and at the angular resolution required to spatially resolve these processes.

\begin{acknowledgements}
The authors would like to thank the entire WISH team for many, many stimulating discussions. HIFI has been designed and built by a consortium of institutes and university departments from across Europe, Canada and the US under the leadership of SRON Netherlands Institute for Space Research, Groningen, The Netherlands with major contributions from Germany, France and the US. Consortium members are: Canada: CSA, U.Waterloo; France: CESR, LAB, LERMA, IRAM; Germany: KOSMA, MPIfR, MPS; Ireland, NUI Maynooth; Italy: ASI, IFSI-INAF, Arcetri-INAF; Netherlands: SRON, TUD; Poland: CAMK, CBK; Spain: Observatorio Astronomico Nacional (IGN), Centro de Astrobiolog{\'i}a (CSIC-INTA); Sweden: Chalmers University of Technology - MC2, RSS \& GARD, Onsala Space Observatory, Swedish National Space Board, Stockholm University - Stockholm Observatory; Switzerland: ETH Z{\"u}rich, FHNW; USA: Caltech, JPL, NHSC. Astrochemistry in Leiden is supported by the Netherlands Research School for Astronomy (NOVA), by a Spinoza grant and grant 614.001.008 from the Netherlands Organisation for Scientific Research (NWO), and by the European Community's Seventh Framework Programme FP7/2007-2013 under grant agreement 238258 (LASSIE).
\end{acknowledgements}

\bibliographystyle{aa}
\bibliography{bibliography}

\appendix

\section{Observation ID's}\label{app:obsid}

All observations presented in this paper and their respective observation IDs are provided in Table \ref{tab:obsid}.

\begin{table*}
\caption{Observation IDs of \textit{Herschel} observations presented and discussed in this paper.}
\begin{center}
\begin{tabular}{l c c c c c c}
\hline \hline
Setting & IRAS2A & IRAS3A & IRAS4A & IRAS4B & SMM1 & SMM3 \\
\hline
H$_2$O 1$_{10}$--1$_{01}$ & 1342202067 & 1342202066 & 1342202065 & 1342202064 & 1342208580 & 1342208579 \\
H$_2$O 2$_{12}$--1$_{01}$ & 1342215966 &  & 1342203951 & 1342203952 & 1342207660 & 1342215965 \\
H$_2$O 1$_{11}$--0$_{00}$ & 1342191657 &  & 1342191656 & 1342191655 & 1342207379 & 1342207377 \\
H$_2$O 2$_{02}$--1$_{11}$ & 1342191606 &  & 1342191605 & 1342191604 & 1342194994 & 1342207658 \\
H$_2$O 2$_{11}$--2$_{02}$ & 1342191748 &  & 1342191749 & 1342191750 & 1342194561 & 1342207617 \\
H$_2$O 3$_{12}$--3$_{03}$ & 1342191654 &  & 1342201796 & 1342201797 & 1342207378 & 1342207376 \\
H$_2$O 3$_{12}$--3$_{03}$\tablefootmark{a} & 1342215968 &  & 1342249014 & 1342249851 & 1342254450 &  \\
H$_2$O 3$_{12}$--2$_{21}$ / CO 10--9\tablefootmark{b} & 1342191701 &  & 1342191721 & 1342191722 & 1342207701 & 1342207699 \\
OH / CO 16--15\tablefootmark{b} &  &  &  &  & 1342208574 &  \\
HCO$^+$ / CH\tablefootmark{c} & 1342192206 &  & 1342192207 & 1342202033 & 1342194463 & 1342207580 \\
CH$^+$\tablefootmark{d} & 1342203229 &  & 1342203230 & 1342203228 & 1342207620 &  \\
OH$^+$\tablefootmark{d} & 1342203180 &  & 1342203181 & 1342203179 & 1342207656 &  \\
C$^+$\tablefootmark{d} & 1342201840 &  & 1342201841 & 1342201842 & 1342208575 &  \\
\hline
\end{tabular}
\end{center}
\tablefoot{
	\tablefoottext{a}{Reobserved as part of OT2 programme OT2\_rvisser\_2.}
	\tablefoottext{b}{Obtained in the same setting.}
	\tablefoottext{c}{Obtained in the same setting as H$_2^{18}$O 1$_{10}$--1$_{01}$.}
	\tablefoottext{d}{Observations to be presented in Benz et al. (in prep.).}
}
\label{tab:obsid}
\end{table*}

\end{document}